\documentclass{webofc}

\usepackage[varg]{txfonts}
\usepackage{hyperref}
\usepackage{url}
\usepackage{physics}
\usepackage{xcolor}
\hypersetup{colorlinks=true,citecolor=blue,urlcolor=blue,linkcolor=blue}

\begin{document}
\title{Initial stage jet momentum broadening in tBLFQ formalism}

\author{\firstname{Dana} \lastname{Avramescu}\inst{1,2}\fnsep\thanks{\email{dana.d.avramescu@jyu.fi}} \and
        \firstname{Carlos} \lastname{Lamas}\inst{3}\fnsep\thanks{\email{carloslamas.rodriguez@usc.es}} \and
        \firstname{Tuomas} \lastname{Lappi}\inst{1, 2}\fnsep\thanks{\email{tuomas.v.v.lappi@jyu.fi}} \and
        \firstname{Meijian} \lastname{Li}\inst{3}\fnsep\thanks{\email{meijian.li@usc.es}} \and
        \firstname{Carlos} \lastname{A. Salgado}\inst{3}\fnsep\thanks{\email{carlos.salgado@usc.es}}
}

\institute{Department of Physics, University of Jyväskylä, P.O. Box 35, 40014 University of Jyväskylä, Finland
\and
           Helsinki Institute of Physics, P.O. Box 64, 00014 University of Helsinki, Finland
\and
           Instituto Galego de Fisica de Altas Enerxias (IGFAE),
            Universidade de Santiago de Compostela, 15705 Galicia, Spain
          }

\abstract{We study the momentum broadening of a high-energy quark jet in the large density gluon medium created right after the collision of two ultrarelativistic heavy nuclei, the Glasma. Previous Glasma studies modeled the jet as a classical probe particle, for which position and momentum are simultaneously determined. In this work, we use the light-front QCD Hamiltonian formalism to treat the jet as a fully quantum state. We compute its real-time evolution while propagating through the Glasma classical background fields, which act as an interaction potential in the quantum evolution of the jet. We present results for the momentum broadening and jet quenching parameter of a jet at mid-rapidity, with special emphasis on the anisotropies  between the longitudinal and transverse directions relative to the collision axis. In addition, we compare our results to classical calculations, and initiate a study of the distinction between kinetic and canonical momentum in the context of jet momentum broadening.
}
\maketitle
\section{Introduction}
\label{intro}
Jets are fundamental probes of heavy-ion collisions. They are created in a hard scattering at the beginning of the process and experience all the stages of the hot and dense nuclear matter created in the collision. The effect of the initial stages was usually neglected, as energy loss was predicted to be suppressed for the first $\sim 0.6\, \mathrm{fm/c}$ after the collision \cite{Andres:2020}. However, studies of jets in Glasma or pre-equilibrium phase showed that those probes can be significantly modified in the initial stages \cite{Ipp:2020, Avramescu:2023qvv, Boguslavski:2023alu, Barata:2024xwy}. These analyses treat the jet as a classical probe particle deflected by the forces from the classical Glasma fields, allowing the calculation of jet momentum broadening during the propagation. However, because jets are inherently quantum objects, a quantum treatment is necessary to properly account for quantum interference and gluon radiation, which cannot be captured by classical calculations.

In this work, we use for the first time a light-front Hamiltonian formalism to compute the real-time evolution of the quantum state associated to the probe particle in the Glasma fields. The modification of the jet wavefunction through the interaction with the Glasma fields enables us to extract various quantities such as medium-induced momentum broadening and energy loss. Here, we focus on studying momentum broadening and leave the investigation of other observables for future works. These results differ from those shown at Hard Probes 2024,  mainly due to a better understanding of the distinction between kinetic and canonical momentum, which we address in Sec~\ref{sec:c_k_momentum}.

\section{The Glasma fields}
\label{glasma}
Following the Color Glass Condensate approach to describe the high-energy nuclei before the collision, an initial condition for the gluon fields right after the collision can be derived~\cite{Kovner:1995}, which, for sufficiently high energies, can be approximated as boost invariant. These fields are then evolved by solving the free Yang-Mills equation $[D_\mu, F^{\mu\nu}] = 0$, yielding the so-called Glasma fields. Solving free Yang-Mills for non-Abelian color fields is non-trivial and there exist several methods. In this work, we numerically evolve the Glasma fields using real-time lattice gauge theory \cite{Krasnitz:1998}. The resulting Glasma fields can be characterized by a single physical scale, the saturation momentum $Q_s$. In this work, we use $Q_s=1.5\,\mathrm{GeV}$, which roughly corresponds to central collisions at the Large Hadron Collider.

\section{Real-time jet evolution}
We consider the propagation of an ultrarelativistic quark jet at mid-rapidity, traveling along the $x$-direction through the Glasma field that is generated by the colliding heavy ions moving along the $z$-direction. We will treat the jet as a fully quantum state and the Glasma as a background gluon field.

\subsection{The time-dependent Basis Light Front Quantization Formalism}
The time-dependent Basis Light-Front Quantization (tBLFQ) approach is a non-perturbative computational method based on the light-front quantum field theory and the Hamiltonian formalism~\cite{tBLFQ}. The tBLFQ formalism for simulating the real-time evolution of a quark jet inside a gluon background field has been developed in Ref.~\cite{tBLFQ:q_jet} and extended to quantum simulation in Ref.~\cite{tBLFQ_QC:q_jet}. There, the jet traverses a SU(3) colored medium described by the MV model, corresponding to the scenarios of cold nuclear matter created in deep inelastic scattering experiments and the quark-gluon plasma produced in ultrarelativistic heavy-ion collisions. In this work, we are interested in the jet evolution in the initial stages of heavy-ion collisions, thereby we take the background gluon field as the Glasma field.

We consider a quark jet interacting with the Glasma background field. As a starting point, we truncate the quark's Fock space to its leading sector, $\ket{q}$. Consequently, the QCD Lagrangian simplifies to the color-Dirac Lagrangian, that is obtained by replacing derivatives with covariant derivatives in the Dirac Lagrangian through the so-called minimal coupling procedure
\begin{align}
    \mathcal{L} = \overline{\Psi} (i\gamma^\mu D_\mu - m)\Psi\, .
    \label{eq:ColorDirac}
\end{align}
Here $D_\mu = \partial_\mu + ig\mathcal{A}_\mu$ and $\mathcal{A}_\mu = \mathcal{A}^a_\mu t^a$ is the classical background color field. There are no quantum gluons in the truncated Fock space, $A^\mu = 0$, thereby we do not consider gluon radiation in the process.

Knowing the Lagrangian, the Hamiltonian can be obtained through a Legendre transform \cite{Brodsky:1998}. In the light-front form, where the light-front time $x^+ = t+x$ plays the role of the temporal variable
\begin{align}
    P^- = \int dx^- d^2\vec{x}_\perp \overline{\Psi} \left[\frac{1}{2} \gamma^+ \frac{m^2-\nabla^{\perp\, 2}}{i\partial^+ -g\mathcal A^+} + g (\gamma^+ \mathcal{A}_+ + \gamma^i \mathcal{A}_i) + \frac{g^2}{2} \gamma^i\mathcal{A}_i \frac{\gamma^+}{i\partial^+-g\mathcal A^+} \gamma^j \mathcal{A}_j \right]\Psi \, ,
    \label{eq:LFHamiltonian}
\end{align}
where $\Psi$ is the quark field. Note that the quantum gauge field,  which must be in light-cone gauge $A^+=0$, is absent here due to the Fock space truncation, whereas the external field $ \mathcal A$ remains in an arbitrary gauge, in our case the Fock-Schwinger gauge, natural to the Glasma fields. We focus on the eikonal limit, where the quark's energy is assumed to be very large, $p^+ \to \infty$, then only the terms not suppressed by $1/p^+$ contribute, reducing the Hamiltonian to
\begin{equation}
    P^- = g \int dx^- d^2\vec{x}_\perp g \bar{\Psi} \gamma^+ \mathcal{A}_+ \Psi\, .
    \label{eq:EikonalHamiltonian}
\end{equation}
Thus only the interaction with the $\mathcal{A}_+$ component of the background fields appears in the eikonal Hamiltonian and all the terms containing $\mathcal{A}_-$ in Eq. \ref{eq:LFHamiltonian} vanish even if we do not impose the gauge condition. The evolution of the jet as a quantum state obeys the time-dependent Schr\"odinger equation on the light front,
\begin{align}
    i \frac{\partial}{\partial x^+} \ket{\psi; x^+} = \frac{1}{2} P^- \ket{\psi; x^+}\;.
    \label{eq:SchrodingerEquation}
\end{align}
We are interested in how the jet's momentum evolve due to interactions with the Glasma background field. Therefore, we choose the basis states $\ket{\beta}$ as the eigenstates of the free QCD Hamiltonian, and expand the jet state as
\begin{equation}
    \ket{\psi; x^+} = \sum_\beta c_\beta(x^+) \ket{\beta}\; ,
    \label{eq:QuarkStateExpansion}
\end{equation}
Each basis state $\ket{\beta}$ is labeled by the quantum numbers of three-momentum, color, and light-front helicity. The light-front wavefunction of the jet is then given in terms of the state vector $\textbf{c}(x^+)$, which consisted of the basis coefficients. The real-time evolution of the quantum state can be expressed as
\begin{equation}
    \textbf{c}(x^+) = \mathcal{T}_+ \exp\left[-i \int_0^{x^+} dt \mathcal{M}(t) \right] \textbf{c}(0)\; ,
    \label{eq:CoefficientEvolution}
\end{equation}
where $\mathcal{M}_{\beta\beta^\prime}(x^+) \equiv \bra{\beta}P^-(x^+)/2\ket{\beta^\prime}$ is the matrix element of the Hamiltonian in the chosen basis space. Once the wavefunction of the jet state is known through $\textbf{c}(x^+)$, observables can be readily evaluated from it.

\subsection{Momentum broadening results}
\begin{figure*}
\centering
\vspace*{1cm}
\includegraphics[width=12cm,clip]{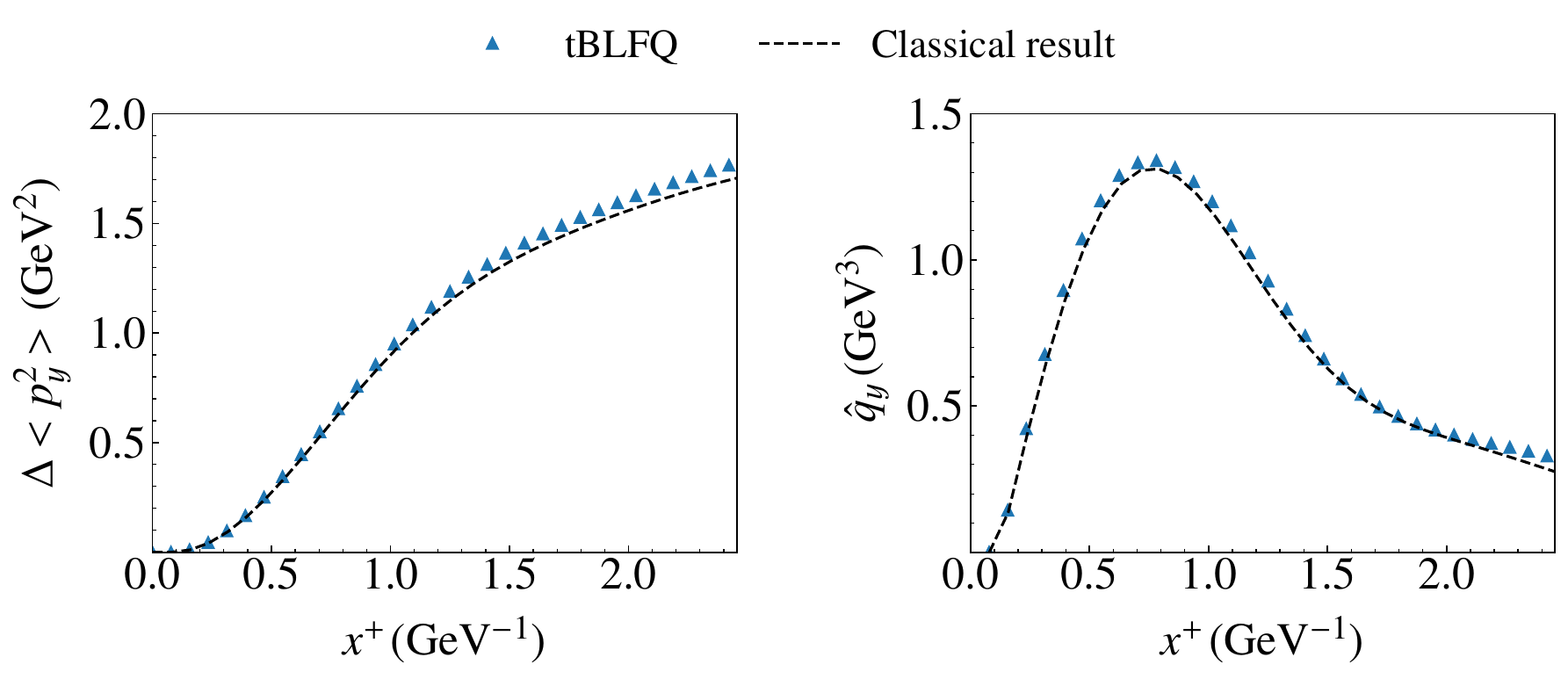}
\caption{Evolution of the expectation value of the jet transverse momentum (on the left panel) and the jet quenching parameter (on the right panel). We define $\Delta \langle p_y^2 (x^+)\rangle = \langle p_y^2 (x^+)\rangle - \langle p_y^2 (0)\rangle$ to subtract the initial width of the wavefunction in momentum space and show only the broadening due to interactions with the medium. The result obtained using tBLFQ formalism is shown in solid triangle and the classical calculation result using Eq. (\ref{eq:ClassicalCanonicMomentum}) is shown in dashed line as a comparison.}
\label{fig:MomentumBroadening}
\end{figure*}

We use the evolved wavefunction in Eq. \ref{eq:CoefficientEvolution} to investigate the momentum broadening effect. To quantify momentum broadening, we extract the expectation value of the transverse momentum square from the jet wavefunction at different times,
\begin{align}
    \expval{p_i^2(x^+)} = \bra{\Psi; x^+} p_i^2 \ket{\Psi; x^+}\, .
    \label{eq:MomentumExpectationValue}
\end{align}
The following results are obtained initializing the jet as a very localized wave package in coordinate space, $x \simeq y \simeq 0$, as this mid-rapidity limit simplifies the calculation. We have however checked that the results are not sensible to the initialization. In Figure~\ref{fig:MomentumBroadening}, we show the results of $ \langle p_y^2(x^+)\rangle$ and the associated quenching parameter $\hat{q}_y = d\langle p^2_y\rangle /dx^+$. Note that the $ y$-direction is perpendicular to the beam direction ($z$-direction). The momentum broadening is negligible along the $z$ direction, as the Glasma field in the early stages is uniform along the collision axis. The simulation results using the tBLFQ formalism are shown in the solid triangle, where we have averaged over multiple configurations of the Glasma fields. In the right panel, the corresponding quenching parameter $\hat{q}_y$ rapidly increases to values much larger than those typical in the Quark-Gluon Plasma \cite{qhat}. It peaks around $x^+ \sim 1/Q_s$. Beyond this point, as the Glasma becomes dilute and less effective at accelerating the parton, $\hat{q}$ decreases and eventually vanishes. For comparison, we also computed the transverse momentum and quenching parameter in the classical formalism. These results are shown in the dashed line in the figure, and they are in a good agreement with the quantum results.

In the classical formalism, the equation of motion for the expectation value of the momentum operator reduces to the classical Hamiltonian equation, $dp^i/dx^+ = \partial p^i/\partial x^+ + \lbrace p^i, H \rbrace_P$, where $\lbrace \ldots \rbrace_P$ represents the Poisson bracket and $H$ is the classical Hamiltonian. In the eikonal limit and imposing the boost invariance of the Glasma fields, $dp^y/dx^+ = g \partial^y \mathcal{A}_x$ and $dp^z/dx^+ = 0$. Momentum broadening in the transverse direction is then obtained by solving
\begin{equation}
    \expval{p_y^2(x^+)}_{\text{config}} = \frac{g^2}{N_c} \int_0^{x^+} dt_1 \int_0^{x^+} dt_2 \expval{\Tr\left[\tilde{f}^y(t_1)\tilde{f}^y(t_2)\right]}_{\text{config}}\;,
    \label{eq:ClassicalCanonicMomentum}
\end{equation}
where $\tilde{f}^y(x^+)=U^\dagger(x^+)f^y U(x^+)$ contains $f^y=\partial^y \mathcal{A}_x$ parallel transported with the Wilson line in the quark direction $U(x^+) = \mathcal{P} \exp\left(-ig\int_0^{x^+} dt \mathcal{A}_x(t)\right)$. Performing averages over color charge configurations yields the factor $1/N_c$. The bracket $ \expval{\ldots}_{\text{config}}$ represents the average over configurations of the Glasma field, and should be distinguished from the expectation value of a quantum state. 

The observed classical–quantum agreement confirms that the quantum expectation value reproduces the classical trajectory. This first application to the early stage of a heavy-ion collision lays the groundwork for future studies of quantum effects like jet energy loss.

\subsection{Canonical and kinetic momentum}\label{sec:c_k_momentum}

\begin{figure}[h]
\centering
\includegraphics[width=8cm,clip]{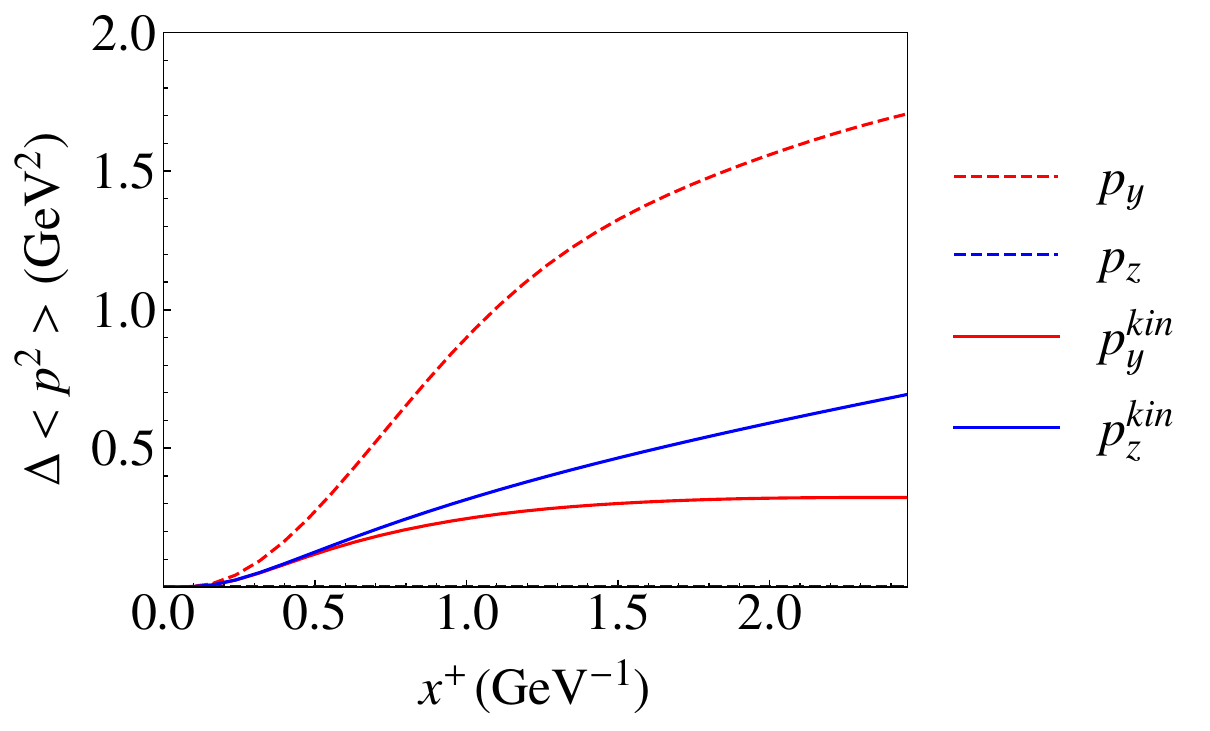}
\caption{Comparison of the canonical and kinetic momenta in the classical formalism. The kinetic (canonical) momentum, shown with solid (dashed) lines, is calculated using Eq.~\eqref{eq:ClassicalKineticMomentum} [Eq.~\eqref{eq:ClassicalCanonicMomentum}].
Momenta in the $y-$ and $z-$ directions are indicated in red and blue, respectively. The $p_z$ line overlaps the horizontal axis, indicating no canonical momentum broadening in $z$.
}
\label{fig:KineticMomentum}
\end{figure}

We would like to note that the canonical momentum conjugated to the position is, in general, not the same as the kinetic momentum, defined as $p^+ d\vec{x}_\perp/dx^+$ in the light-front form. For a particle propagating in an external field $\vec{p}_{kin} = \vec{p} - g\vec{\mathcal{A}} \neq \vec{p}$. In the preceding section, we studied jet broadening via the canonical momentum and found a good quantum–classical agreement. We now extend the classical analysis to kinetic momentum, extracted as~\cite{Ipp:2020}, 
\begin{equation}
    \expval{p_{\mathrm{kin},i}^2(x^+)}_{\text{config}} = \frac{g^2}{N_c} \int_0^{x^+} dt_1 \int_0^{x^+} dt_2 \expval{\Tr\left[\widetilde{F}^i(t_1)\widetilde{F}^i(t_2)\right]}_{\text{config}}\, ,
    \label{eq:ClassicalKineticMomentum}
\end{equation}
where $\widetilde{F}^i(x^+)=U^\dagger(x^+)F^i U(x^+)$ for $i=y,z$, and $F^y=E^y-B^z$ and $F^z=E^z+B^y$ contains the Glasma electric and magnetic fields. Figure \ref{fig:KineticMomentum} shows that the classical simulation predicts larger broadening in the beam direction ($z$-direction) than in the transverse direction ($y$-direction) for the kinetic momentum, contrasting with the canonical momentum.

Though in actual measurement it is the kinetic momentum being observed, in high-energy experiments, where the final particles of the jets are detected, the background field generated by the collision is already washed out and the kinetic and canonical momentum are equivalent. In our theoretical calculations, we have the ability to access the intermediate states, and it would be interesting to also extract the kinetic momentum from the quantum simulation, which we leave for further investigation.

\section{Summary}
We have, for the first time, computed the real-time quantum evolution of a quark jet interacting with the initial-stage Glasma fields. Using the obtained jet light-front wavefunction, we extracted the expectation value of the transverse momentum square and the jet quenching parameter. Our results show good agreement with the classical calculations when expressed in terms of the canonical momentum. Additionally, we initiated a study of the distinction between the kinetic and the canonical momentum in the context of jet momentum broadening.

\section*{Acknowledgement}
\vspace{-.1cm}
DA and TL are supported by the Research Council of Finland, the Centre of Excellence in Quark Matter (projects 346324 and 364191). DA also acknowledges the support of the Vilho, Yrj\"{o} and Kalle V\"{a}is\"{a}l\"{a} Foundation. CL, ML, and CS are supported by the European Research Council under project ERC-2018-ADG-835105 YoctoLHC; by Maria de Maeztu excellence unit grant CEX2023-001318-M and project PID2020-119632GB-I00 funded by MICIU/AEI/10.13039/501100011033; and by ERDF/EU. 
ML also acknowledges the support of Xunta de Galicia under the ED431F 2023/10 project.

\end{document}